\documentclass[sigconf,authorversion,nonacm]{acmart}
\usepackage[utf8]{inputenc}
\usepackage{bm}
\usepackage{balance}
\usepackage{bbding}
\usepackage{booktabs} % For formal tables
\usepackage{graphics}
\usepackage{multirow}
\usepackage{array}
\usepackage{flushend}
\usepackage[english]{babel}
\usepackage{textcomp}
\usepackage{latexsym}
\usepackage{natbib}
\usepackage{graphicx}
\usepackage{multirow}
\usepackage[labelformat=simple]{subcaption}
\usepackage{color}
\usepackage{dsfont}
\usepackage{algorithm}
\usepackage{algorithmic}
\usepackage{tabularx}

\newcommand{\ie}{\textit{i.e.}}
\newcommand{\eg}{\textit{e.g.}}
\usepackage{algorithm}
\usepackage{algorithmic}
\begin{document}
\settopmatter{printacmref=false, printfolios=false}
% \newcommand{\zhou}[1]{\textcolor{blue}{[[@zhou: #1]]}}

% \copyrightyear{2020} 
% \acmYear{2020} 
% \setcopyright{acmlicensed}\acmConference[SIGIR '20]{Proceedings of the 43rd International ACM SIGIR Conference on Research and Development in Information Retrieval}{July 25--30, 2020}{Virtual Event, China}
% \acmBooktitle{Proceedings of the 43rd International ACM SIGIR Conference on Research and Development in Information Retrieval (SIGIR '20), July 25--30, 2020, Virtual Event, China}
% \acmPrice{15.00}
% \acmDOI{10.1145/3397271.3401175}
% \acmISBN{978-1-4503-8016-4/20/07}

\title{DemoRank: Selecting Effective Demonstrations for Large Language Models in Ranking Task}

\author{Wenhan Liu}
\affiliation{Gaoling School of Artificial Intelligence\\
  \institution{Renmin University of China}
  \city{Beijing}
  \country{China}
}
\email{lwh@ruc.edu.cn}

\author{Yutao Zhu}
\affiliation{Gaoling School of Artificial Intelligence\\
  \institution{Renmin University of China}
  \city{Beijing}
  \country{China}
}
\email{yutaozhu94@gmail.com}

\author{Zhicheng Dou}
\authornote{Zhicheng Dou is the corresponding author.}
\affiliation{Gaoling School of Artificial Intelligence\\
  \institution{Renmin University of China}
  \city{Beijing}
  \country{China}
}
\email{dou@ruc.edu.cn}

% \author{Wenhan Liu, Yutao Zhu, \and Zhicheng Dou$^{\dag}$ \\
% Gaoling School of Artificial Intelligence, Renmin University of China \\
% \texttt{lwh@ruc.edu.cn, yutaozhu94@gmail.com, dou@ruc.edu.cn}
% }

\begin{abstract}
Recently, there has been increasing interest in applying large language models (LLMs) as zero-shot passage rankers. However, few studies have explored how to select appropriate in-context demonstrations for the passage ranking task, which is the focus of this paper. Previous studies mainly use LLM's feedback to train a retriever for demonstration selection. These studies apply the LLM to score each demonstration independently, which ignores the dependencies between demonstrations (especially important in ranking task), leading to inferior performance of top-$k$ retrieved demonstrations. To mitigate this issue, we introduce a demonstration reranker to rerank the retrieved demonstrations so that top-$k$ ranked ones are more suitable for ICL. However, generating training data for such reranker is quite challenging. On the one hand, different from demonstration retriever, the training samples of reranker need to incorporate demonstration dependencies. On the other hand, obtaining the gold ranking from the retrieved demonstrations is an NP-hard problem, which is hard to implement. To overcome these challenges, we propose a method to approximate the optimal demonstration list iteratively and utilize LLM to score demonstration lists of varying lengths. By doing so, the search space is greatly reduced and demonstration dependencies are considered. Based on these scored demonstration lists, we further design a list-pairwise training approach which compares a pair of lists that only differ in the last demonstration, to teach the reranker how to select the next demonstration given a previous sequence. In this paper, we propose a demonstration selection framework DemoRank for ranking task and conduct extensive experiments to prove its strong ability.

% Previous studies mainly apply a demonstration retriever to retrieve demonstrations and use top-$k$ demonstrations for in-context learning (ICL). We claim that this approach overlooks the dependencies between demonstrations, leading to inferior performance of few-shot demonstrations in the passage ranking task. In this paper, we propose to use a demonstration reranker to rerank the retrieved demonstrations so that the top-$k$ reranked ones will be more suitable for ICL. However, generating training data for such reranker is quite challenging. On the one hand, 

% In this paper, we formulate the demonstration selection as a \textit{retrieve-then-rerank} process and introduce the DemoRank framework for demonstration selection. In this framework, the demonstration retriever is trained using individually scored demonstrations, while the demonstration reranker is trained based on dependency-aware training samples. We propose an efficient method to iteratively construct these samples and design a list-pairwise training approach to optimize the reranker. Extensive experiments on a series of passage ranking datasets prove the superior performance of DemoRank. Further analysis shows the effectiveness of each proposed component and DemoRank's powerful ability in various scenarios.
\end{abstract}

% <concept_id>10002951.10003317.10003331.10003271</concept_id>
% <concept_desc>Information systems~Personalization</concept_desc>

% \ccsdesc[500]{Information systems~Personalization}

\keywords{Passage Ranking, Large Language Model, In-context Learning}

\def\authors{Wenhan Liu, Yutao Zhu, Zhicheng Dou}

\maketitle

% \begin{figure}[!t]
% 	\centering
% 	\vspace{-0.2cm}
% 	\setlength{\abovecaptionskip}{0.1cm}
% %\   \setlength{\belowcaptionskip}{0.1cm}
% 	\includegraphics[width=1\linewidth]{example.pdf}
% 	\caption{An example for query clarification.\textcolor{red}{to check}}
% 	\label{fig:example}
% \end{figure}

\section{Introduction}
\label{sec:intro}
Large language models (LLM) have demonstrated remarkable performance across a spectrum of natural language processing (NLP) tasks. Recently, there has been significant interest in using LLMs for passage ranking tasks~\cite{beyond_yes_no, rankgpt, qin2023large}. A typical approach is relevance generation, which judges the relevance of a query-passage pair in a pointwise manner. This method prompts LLMs to assess the relevance of a passage to a query by generating responses such as ``Yes'' or ``No''. The relevance score is then computed based on the log-likelihood of these responses. This approach has been demonstrated to be effective in previous studies~\cite{beyond_yes_no, liang2022holistic}.

\begin{figure}[t]
	\centering
	\includegraphics[width=1\linewidth]{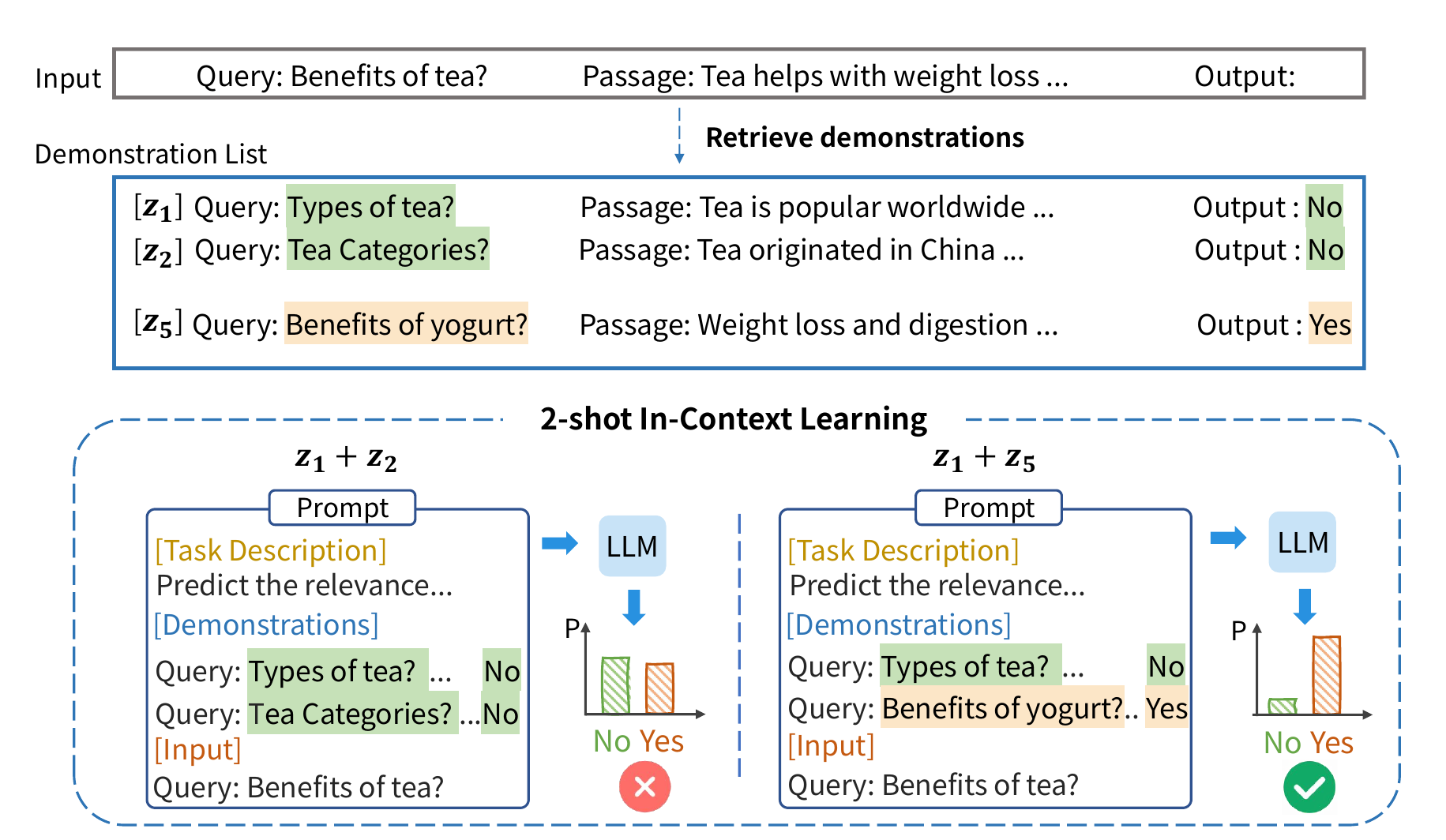}
	\caption{Compared with choosing top-2 demonstrations ($z_1$ and $z_2$), the combination of $z_1$ and $z_5$ provides richer and more diverse query-passage relationships, thus yielding better relevance assessment.}
	\label{fig:intro_example}
       \vspace{3mm}
\end{figure}

In-context learning (ICL) has been proved as an emergent ability of LLMs~\cite{emergent}, enabling them to adapt to specific tasks through several task demonstrations (\ie, input-output examples). This mechanism facilitates the LLMs' comprehension of both the objective and the expected output format of the task, thereby enhancing performance. Many studies have investigated the optimal selection of demonstrations for NLP tasks~\cite{LuBM0S22, ZhangFT22, udr, llmr, D2RCU}, highlighting the importance of tailored demonstrations in achieving high performance. However, the application of ICL to passage ranking tasks has not been extensively studied. Given the complex nature of passage ranking, ICL presents a challenging yet promising opportunity to enhance LLMs' performance. Consequently, this study aims to develop effective demonstration selection strategies to optimize the application of ICL in passage ranking.

A widely-used and effective approach for demonstration selection is training a demonstration retriever using LLM's feedback~\cite{llmr, RubinHB22, udr, UPRISE, RetICL, Dr_icl}. This approach first utilizes an LLM to score some demonstration candidates based on LLM's likelihood of producing the correct output given each candidate and the input, and choose positive and negative candidates based on scores for retriever training. Following this technique line, we propose to train a demonstration retriever tailored for passage ranking task based on LLM's feedback.

In the inference stage, a common practice~\cite{llmr} is to use the trained retriever to retrieve a list of demonstrations and concatenate the top-retrieved ones together in the prompt for ICL. Despite its effectiveness in NLP tasks, directly extending it into the passage ranking task may result in sub-optimal performance. The main challenge lies in the complex nature of the query-passage relationship in passage ranking, which may require a \textit{combination} of multiple demonstrations to provide effective information for understanding such a relationship. Figure~\ref{fig:intro_example} shows an example of such problem. When selecting a 2-shot demonstration for the current input (a relevant query-passage pair), existing methods~\cite{llmr, RubinHB22} will choose the top-2 demonstrations ($z_1$ and $z_2$) returned by the retriever. However, we deem that combining $z_1$ and $z_5$ is more suitable for this case. This is because $z_1$ and $z_5$ have more distinct queries and opposite outputs (relevance label), which provide LLM with \textit{richer and more diverse query-passage relationship signals}, thus contributing more to the relevance assessment. This example shows the insufficiency of pure relevance-based demonstration selection in the few-shot LLM-based passage ranking task. In this paper, we transform the problem of selecting the optimal $k$-shot demonstration from retrieved $n$ demonstrations into a demonstration reranking problem and propose to train a dependency-aware demonstration reranker that iteratively selects the next demonstration given the previous demonstration sequence.

Nevertheless, training such a reranker is a very challenging task. As previously mentioned, it is unreasonable to use LLM's feedback on each individual demonstration for training a reranker designed for $k$-shot selection, because demonstrations can influence each other. Additionally, constructing the ground truth ranking of a reranker tailored for $k$-shot selection requires finding the optimal $k$-shot permutation from the retrieved $n$ demonstrations, which is an NP-hard problem. Theoretically, this requires using LLM to score total $\frac{n!}{(n-k)!}$ demonstration permutations, which is highly time-consuming and impractical. To overcome these challenges, we propose an efficient approach to construct a kind of dependency-aware training samples. Specifically, our approach approximates the optimal demonstration list by iteratively selecting demonstrations from retrieved ones, which significantly reduces the search space. In each iteration, we use an LLM to score a set of demonstration lists that differ only in the last demonstration (\eg, ${[z_3, z_1, z_2], [z_3, z_1, z_4], [z_3, z_1, z_5], \dots}$). Taking the whole list as input, the LLM score considers the dependencies among demonstrations. After constructing the training samples, we design a novel list-pairwise training method which compares a pair of demonstration lists (e.g., $([z_3, z_1, z_2], [z_3, z_1, z_4])$), so that the reranker learns to select the next demonstration given a previous sequence.

% construct a kind of dependency-aware training samples (a list of demonstrations with ranking labels) for reranker training. Specifically, given a retrieved demonstration set, we greedily select demonstrations from the set and annotate them with different ranking labels (from highest to lowest). Each time, the demonstration that maximizes the LLM's feedback when concatenated with the already selected ones is chosen. This process not only considers the dependencies between current demonstration and previously selected ones, but also greatly reduces the number of LLM inferences.

To this end, we propose DemoRank, a \textbf{Demo}nstration selection framework for passage \textbf{Rank}ing, using a two-stage ``retrieve-then-rerank'' strategy. In this framework, we first introduce a demonstration retriever DRetriever trained using demonstrations individually scored by LLM. Then, we introduce a dependency-aware demonstration reranker DReranker to iteratively select demonstrations, so as to obtain a reranked demonstration list. To address the challenges of its training, we propose a method to construct dependency-aware training samples that not only incorporates demonstration dependency but is also time-efficient. We also propose a list-pairwise approach for DReranker training, which focuses on comparing a pair of demonstration lists.

Experiments on a series of ranking datasets prove the effectiveness of DemoRank, especially in few-shot ICL. Further analysis also demonstrates the contribution of each proposed component and DemoRank's strong ability under different scenarios, including limited training data, different demonstration numbers, unseen datasets, different LLM rankers, etc.

The main contributions of our paper are summarized as follows:

(1) To the best of our knowledge, we are the first to comprehensively discuss effective demonstration selection in passage ranking and propose DemoRank which consists of a demonstration retriever and a dependency-aware demonstration reranker.

(2) We propose an efficient method for constructing dependency-aware training samples for demonstration reranker.

(3) Based on these training samples, we design a list-pairwise training approach to teach the reranker to select the next demonstration based on some contexts. 
% To train the demonstration reranker, we design an efficient method for constructing dependency-aware training samples and introduce a novel list-pairwise training approach for optimization.

% (3) Extensive experiments prove significant performance improvements of DemoRank, highlighting its ability to select effective few-shot demonstrations for ranking task.

\section{Related Work}
\subsection{LLM for Passage Ranking}
With the development of large language models (LLMs) in information retrieval~\cite{llm4ir_survey}, there have been many studies exploring how to utilize LLMs for the passage ranking task. In general, these studies can be divided into three categories: pointwise~\cite{liang2022holistic, sachan2022improving}, pairwise~\cite{qin2023large}, and listwise methods~\cite{rankgpt}. Pointwise methods assess the relevance between a query and a single passage. A typical approach is relevance generation~\cite{liang2022holistic, beyond_yes_no}, which provides LLM with a query-passage pair and instructs it to output ``Yes'' if the passage is relevant to the query or ``No'' if not. The relevance score can be calculated based on the generation probability of the token ``Yes''. Another approach of pointwise methods is query generation~\cite{sachan2022improving, opensource}, which calculates relevance score based on the log-likelihood of generating the query based on the passage. Pairwise methods compare two passages at a time and determine their relative relevance to a query, and listwise methods directly rank a passage list. 

Despite promising results, these studies only focus on the zero-shot scenarios, with less emphasis on how to select effective few-shot demonstrations. Manually written or rule-based selection~\cite{parade} is inflexible for ranking tasks. In this paper, we explore more effective few-shot demonstrations selection approaches for ranking tasks. Previous studies~\cite{inters} have revealed that relevance generation of the pointwise method is the most suitable method for passage ranking on open-source LLMs compared with other methods. Thus, we intend to use the relevance generation approach for passage ranking in this paper.

\subsection{Demonstration Retrieval}
A widely used demonstration selection approach is demonstration retrieval. Prior studies have explored using different retrievers for demonstration retrieval, which can be divided into two categories. One is utilizing off-the-shelf retrievers. For example, \citet{AgrawalZLZG23} propose to use BM25 for demonstration retrieval in machine translation. \citet{LiuSZDCC22} propose to use a kNN-based retriever to retrieve demonstrations semantically similar to the test input. The other is to train a demonstration retriever using task-specific signals. For example, \citet{RubinHB22} propose to distill the LLM's feedback to a dense retriever EPR for the semantic parsing task. ~\citet{UPRISE} focuses on training a demonstration retriever that improves LLMs' performance in the cross-task and cross-model scenarios. \citet{udr} and \citet{llmr} propose to train the retriever iteratively on various NLP tasks. However, a common issue with these methods is that they directly choose the top-retrieved demonstrations, which may include redundant information and contribute little to the LLM's understanding of relevance. In this paper, we take the demonstration dependencies into account and introduce a framework that first retrieves a list of demonstrations and then reranks in a dependency-aware manner, better aligning with the few-shot ICL in the ranking task.

\section{Preliminaries}
\subsection{Relevance Generation for Ranking Task}
Passage ranking aims to rank a list of retrieved passages based on their relevance to a query. Formally, given a query $q$ and a passage list $\left[p_1, \dots, p_n\right]$, our task is to compute a relevance score $S(q, p_i)$ for each passage. In the LLM-based relevance generation methods~\cite{liang2022holistic, beyond_yes_no}, an LLM is provided with a prompt consisting of a query and a passage, and instructed to output a binary label ``Yes'' or ``No'' to indicate whether the passage is relevant to the query or not. Then a softmax function is applied to the logits of tokens ``Yes'' and ``No'', and the probability of the token ``Yes'' is used as the relevance score:
\begin{equation}
\label{eq:score}
    Rs(q, p_i) = \text{Pr}(\text{``Yes''}|T, q, p_i),
\end{equation}
where $T$ is the task description. Finally, the passages are ranked based on the relevance score $S(q, p_i)$ in descending order.

\subsection{In-context Learning in Ranking Task}
In-context learning is a technique that inserts a few demonstrations into the prompt to help LLMs perform a task without updating parameters. In relevance generation task, given $k$ in-context demonstrations $\{z_i\}_{i=1}^k$, where $z_i=(\hat{q},\hat{p},\hat{y})$ is a triple consisting of a query, a passage and a binary output (``Yes'' or ``No'') indicating the relevance label, the relevance score $Rs(q, p_i)$ is calculated by:
\begin{equation}
    Rs(q, p_i) = \text{Pr}(\text{``Yes''}|T, \{z_i\}_{i=1}^k, q, p_i),
\end{equation}
where $T$ is the task description, which is used in ICL to help LLMs understand the task~\cite{inters,udr}.

\begin{figure*}[!tb]
  \centering
  \vspace{-0.4cm} % 增加图形上方的空白以往下移图形
  \setlength{\abovecaptionskip}{-0.1cm} % 缩小标题和图之间的距离
  \includegraphics[width=1\linewidth]{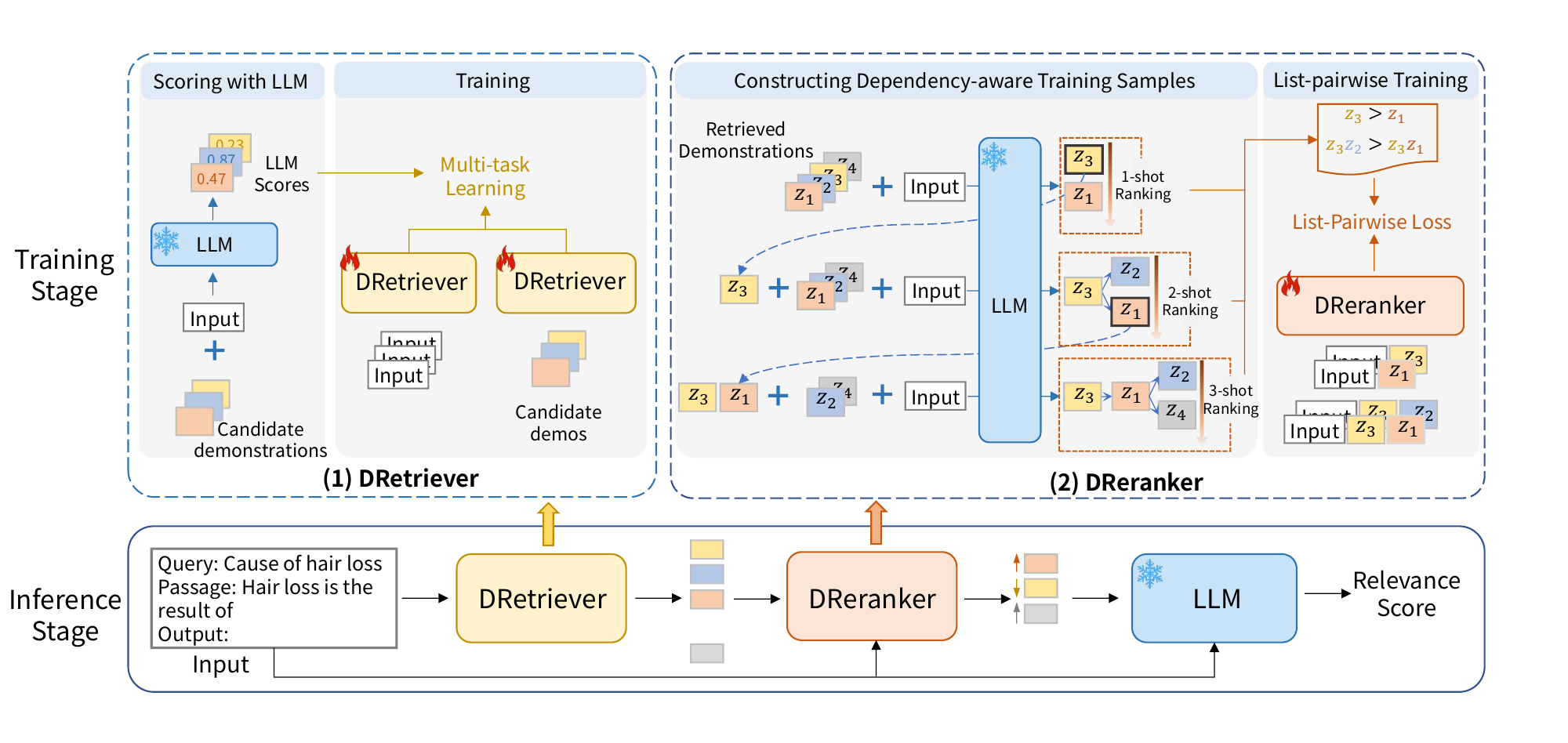}
  \caption{An overview of our proposed framework DemoRank. It comprises two main components: DRetriever and DReranker. We first train the DRetriever using demonstration candidates individually scored by LLM. Then, We iteratively construct a set of dependency-aware training samples (each sample is a ranking of a $k$-shot demonstration list) and use a list-pairwise training approach to optimize the DReranker. During inference, we first use DRetriever to obtain a list of demonstrations and then apply DReranker to iteratively select demonstrations to obtain the top-$k$ reranked demonstrations for ICL.}
  \label{fig:model}
  % \vspace{-2.5cm}
\end{figure*}

\section{The DemoRank Framework}
As shown in Figure~\ref{fig:model}, our DemoRank framework follows a process of demonstration retrieval and dependency-aware reranking. We construct the training data for demonstration retriever DRetriever by individually scoring each demonstration candidate. As for the demonstration reranker DReranker, as we mentioned in Section~\ref{sec:intro}, it is unreasonable to individually score each demonstration and impractical to obtain the optimal demonstration ranking. To overcome these challenges, we construct a kind of dependency-aware training samples in an efficient way for DReranker training. Based on these training samples, we further propose a list-pairwise training approach for model optimization. In this section, we will elaborate on our demonstration pool construction, the pipeline of model training, and inference.

\subsection{Demonstration Pool Construction.} \label{demo_pool}
Given a passage ranking dataset (\eg, MS MARCO~\cite{MSMARCO}), we use its training set to construct our demonstration pool $\mathcal{P}$. For each query in the training set, we construct positive and negative demonstrations by pairing the query with its relevant and irrelevant passages respectively. To maintain the output label balance in the demonstration pool $\mathcal{P}$, the number of negative demonstrations of each query is set equal to its positive demonstrations.

% The relevant passages are those that have been annotated for the query in the dataset and the irrelevant passages are sampled from the top-100 passages retrieved by BM25.

\subsection{Demonstration Retriever DRetriever} \label{sec:retriever}
In this part, we train DRetriever to retrieve potentially useful demonstrations for subsequent dependency-aware reranking. We apply an LLM to score a set of demonstration candidates to obtain supervised signals and use them to train the retriever through a multi-task learning strategy.

\subsubsection{Scoring with LLM}
For a training input $I = (q, p)$ which contains a query-passage pair, we select a set of demonstrations from demonstration pool $\mathcal{P}$ as training candidates. Following previous studies~\cite{llmr}, we employ the BM25 algorithm to retrieve top-$b$ demonstrations. Due to the complex nature of passage ranking, the utility of a demonstration is not directly related to its similarity to the input~\cite{parade}. To include more potential useful demonstrations for training, we also randomly sample another $b$ demonstrations from $\mathcal{P}$. The total number of training candidates is annotated as $N$ ($N=2*b$). 

After that, we apply a frozen LLM scorer to score each demonstration $z_i$ for the training input $I$ using the following equation:
\begin{equation} \label{eq:llm_score}
    f(z_i, I) = \frac{\text{Pr}(y|T, z_i, I)}{\sum_{y' \in Y}\text{Pr}(y'|T, z_i, I)},
\end{equation}
where $y$ is the relevance label for the query-passage pair in $I$, $Y=\{``\text{Yes}", ``\text{No}"\}$ is the label space and $T$ is the task description. In this paper, the LLM scorer uses the same model as the LLM passage ranker. Nevertheless, we also explored the transferability of LLM scorer on different LLM passage rankers in Section~\ref{subsubsec:transferability}).

\subsubsection{Training} Our DRetriever is based on bi-encoder architecture. Given the current training input $I=(q, p)$ and a candidate $z_i$, we use encoder $E_I$ and demonstration encoder $E_z$ to encode them respectively and calculate the similarity score as:
\begin{equation}
    S(I, z_i) = E_I(I)^\top E_z(z_i),
\end{equation}
where the two encoders $E_I$ and $E_z$ share same parameters and encode with average pooling.

Then we apply a contrastive loss $L_\text{c}$ to maximize the score between the training input $I$ and positive demonstration $z^+$ and minimize it for negative demonstration $z_i^-$. Here $z^+$ is the demonstration with the highest LLM score and $z_i^-$ are the remaining ones. The contrastive loss $L_\text{c}$ is calculated as:
\begin{equation}
    L_\text{c} = -\log \frac{e^{S(I, z^+)}}
    {\sum_{z' \in Z} e^{S(I, z')}},
\end{equation}
where $Z=\{z^+, z_1^-, \dots, z_{N-1}^- \}$. Here we choose not to use in-batch negatives. The reasons are discussed in Section~\ref{subsubsec:variants}.

To make use of the fine-grained supervision of LLM's feedback, we also consider a ranking loss RankNet~\cite{ranknet} to inject the ranking signal of candidates into training:
\begin{equation} \label{eq:L_r}
    L_\text{r} = \sum_{i,j}^{|Z|} \mathds{1}_{r_i < r_j} \log(1 + e^{S(I,z_j)-S(I, z_i)}),
\end{equation}
where $r_i$ is the rank of $z_i$ in $Z$ when sorted in descending order by the LLM score.

The final loss function $L$ is defined as the weighted sum of contrastive loss $L_{c}$ and ranking loss $L_{r}$:
\begin{equation}
    L=\lambda L_{c} + L_{r},
\end{equation}
where $\lambda$ is a pre-defined hyper-parameter.

\subsection{Demonstration Reranker DReranker} \label{subsec:reranker}
Previous studies~\cite{llmr, RubinHB22, udr} mainly use the top-$k$ retrieved demonstrations for ICL which ignores the demonstration dependencies and could be sub-optimal for ranking tasks. In this part, we introduce a novel demonstration reranker DReranker to rerank the retrieved demonstrations to make the top-$k$ more suitable for $k$-shot ICL. As mentioned in Section~\ref{sec:intro}, generating training data for the DReranker presents significant challenges: (1) LLM's feedback on individual demonstrations used for retriever training fails to capture the demonstration dependencies; (2) constructing the ground truth ranking for $k$-shot demonstration requires using LLM scorer to evaluate all possible permutations, which has high time complexity and impossible. To overcome these challenges, we design a time-efficient approach to iteratively approximate the optimal demonstration sequence and generate a series of scored demonstration lists at each iteration, which will be used as the training samples for DReranker. In this process, the LLM scorer evaluates multiple demonstrations at a time, thus taking the demonstration dependencies into account. Finally, we propose a list-pairwise training approach which compares a pair of demonstration lists (with only the last demonstration being different). This approach teaches the DReranker how to select the next demonstration given the previously selected ones.

% To mitigate this issue, we propose to train a demonstration reranker DReranker, which iteratively selects demonstrations from the retrieved set to obtain a reranked demonstration list. In each iteration, DReranker selects a demonstration by considering its dependency with previously selected ones. To train DReranker, we design an efficient approach for constructing a kind of dependency-aware training samples. Specifically, we iteratively select demonstration, which yield high LLM score when concatanated with selected demonstrations, to approximate the optimal demonstration list. In the $k$-th iteration, we concatenate each unselected demonstration with the selected demonstration sequence respectively, use an LLM to score and generate a ranked list of $k$-shot demonstration (e.g., ${[z_3, z_1, z_2], [z_3, z_1, z_4], \dots}$ for 3-shot Ranked List in Figure~\ref{fig:model}). Here, the LLM scores multiple demonstrations (\eg, $z_3$, $z_1$ and $z_2$) at a time, thereby considering the demonstration dependency, which is different from the training data of demonstration retriever. Finally, we propose a list-pairwise loss for DReranker optimization, which helps the DReranker learn to construct a good demonstration list by comparing a pair of demonstration list which differ only in the last demonstration.

\begin{algorithm}[htb]
%  \SetAlgoNoLine  %去掉之前的竖线
\caption{Constructing dependency-aware training samples}
\label{ranking_data_construction}
\begin{algorithmic}[1] %此处的[1]控制一下算法中的每句前面都有标号
\REQUIRE Training input $I$, maximum iteration $K$.
\ENSURE Dependency-aware training samples $O$.
\STATE Retrieve top-$M$ demonstrations $Z^\text{r}$
\STATE $O \leftarrow \{\}$, selected demonstrations $Z^s \leftarrow []$, unselected demonstrations $Z^u \leftarrow Z^\text{r}$
\FOR{$i = 1$ to $K$}
    \STATE Score and rank each $z_j \in Z^u$ using $f(Z^s \oplus z_j, I)$ based on Equation~(\ref{eq:llm_score})
    \STATE Append $(I,Z^s,Z^u)$ and the ranking of scored $Z^u$ to $O$
    \STATE Select a demonstration $z^* \in Z^u$ based on Equation~(\ref{eq:sample}) \\
    \STATE Append $z^*$ to $Z^s$, remove $z^*$ from $Z^u$
    % \STATE $S \leftarrow \left[S, z^* \right]$, $Y \leftarrow Y \cup \{(z^*,y)\}$
\ENDFOR
\RETURN $O$
\end{algorithmic}
\end{algorithm}

\subsubsection{Constructing Dependency-aware Training Samples.} As we mentioned above, our Dreranker acts as a sequential demonstration selector. To generate its training data, we propose to approximate the optimal demonstration sequence iteratively and construct a kind of dependency-aware training samples. Specifically, given a training input $I$, we use our trained DRetriever to retrieve top-$M$ demonstrations $Z^\text{r}$ from the demonstration pool. For the first iteration, we use the LLM scorer to score and rank these demonstrations based on Equation~(\ref{eq:llm_score}). Then, we design a sampling strategy to select a demonstration from the unselected demonstrations $Z^\text{u}$ ($Z^\text{u}$=$Z^\text{r}$ for the first iteration) based on its rank:
\begin{equation}
\label{eq:sample}
    p(z_i) = \frac{g(z_i)}{\sum_{z_j \in Z^\text{u}} g(z_j)}, \quad g(z_i) = \text{exp}(-r_i),
\end{equation}
where $r_i$ is the rank of demonstration $z_i$. This strategy ensures that higher-scoring demonstrations are more likely to be selected for constructing the optimal demonstration sequence. Meanwhile, it retains the possibility of selecting lower-scoring demonstrations, which helps DReranker select the next demonstration when the previous sequence is not good enough, thus improving its ability to handle previous sequences of different quality. For the $k$-th iteration, we use the LLM scorer to score each unselected demonstration by concatenating it with previously selected demonstration sequence and obtain a ranked list. Then we select a demonstration based on the same sampling strategy. Meanwhile, we obtain a ranking of $k$-shot demonstration list (\eg, 3-shot ranking $([z_3, z_1, z_2] > [z_3, z_1, z_4] > \dots)$ in Figure~\ref{fig:model}), which acts as a dependency-aware training sample and will be used by our list-pairwise training approach in the next section. In this way, the dependency between each demonstration and previous demonstrations is considered. Note that as the number of iterations increases, the computational cost of LLM inference also increases. Due to limited computational resources, we set a maximum iteration number $K$. Algorithm~\ref{ranking_data_construction} shows the whole constructing process.

% concatenate the  with each unselected demonstration respectively, score them using an LLM and obtain a ranked list of $k$-shot demonstration (e.g., 3-shot Ranked List ${[[z_3, z_1, z_2], [z_3, z_1, z_4], \dots]}$ in Figure~\ref{fig:model}). Instead of evaluating a single demonstration, the LLM scores multiple demonstrations (\eg, $z_3$, $z_1$ and $z_2$) at once, thus taking the demonstration dependency into account. Then we 

% Then, we iteratively select demonstrations from $Z^\text{r}$ and annotate each of them with a ranking label, as Figure~\ref{fig:model} shows. In each iteration, we select, from the unselected demonstrations in $Z^\text{r}$, the one that maximizes the LLM's feedback when concatenated with already selected ones. Once a demonstration is selected, we append it to the training samples. This process considers previous demonstration sequence when selecting the current demonstration and approximates the optimal $k$-shot demonstration permutation incrementally, which is time-efficient and aligns with the few-shot setting. Note that as the number of iterations increases, the computational cost of LLM inference also increases. Due to limited computational resources, we set a maximum iteration number $K$. After the $K$-th iteration is completed, we annotate a ranking label from $K$ to 1 to each demonstration in the training sample according to their selection order and annotate 0 to the unselected demonstrations in $Z^\text{r}$. Algorithm~\ref{ranking_data_construction} shows this procedure.

\subsubsection{List-pairwise Training}
Learning to rank approaches are usually divided into three categories: pointwise, pairwise and listwise~\cite{liu2009learning}. Demonstration reranking is naturally a listwise problem because the score of a demonstration depends on previous demonstrations. Take the 3-shot demonstration in Figure~\ref{fig:model} as an example, given that $z_3$ and $z_1$ have been selected, $z_2$ have a higher LLM score than $z_4$. In this part, we propose a list-pairwise training method to optimize our DReranker. We call it list-pairwise because the loss is calculated by comparing a pair of demonstration lists ($l_1, l_2$) where $l_1$ and $l_2$ differ only in the last demonstration. With the dependency-aware training samples from 1-shot to $K$-shot already constructed, we generate all pairs of demonstration lists for each shot (for 1-shot, the list length is 1). Then, we calculate our list-pairwise loss similarly to the pairwise RankNet loss:
\begin{equation}
    L_\text{list-pairwise} = \sum_{k=1}^{K} \sum_{i,j}^{|L_k|} \mathds {1}_{r_i < r_j} \log \left(1 + e^{\text{score}(I,l_j) - \text{score}(I,l_i)} \right),
\end{equation}
where $L_k$ is the ranked list of $k$-shot demonstration list, $r_i$ is the rank of $i$-th demonstration list $l_i$ in $L_k$. Our DReranker is based on a cross-encoder model, which calculates $\text{score}(I, l_i)$ by taking the concatenation of the training input $I$ and demonstration list $l_i$ as input, and output a score using the representation of ``[CLS]'' token. Observing demonstration lists of different shots helps the DReranker learn to select the optimal demonstration under a given previous demonstration sequence.

% After constructing the dependency-aware training sample, we obtain a ranking label for each demonstration candidate in $Z^\text{r}$. We employ a cross-encoder model to train our DReranker. The model takes as input the concatenation of training input $I$ and one candidate $z_i$ with a ``[SEP]'' token and outputs a prediction score $s_i$ using the representation of ``[CLS]'' token. Then we apply the RankNet loss function to optimize the reranker model, similar to Equation~(\ref{eq:L_r}):
% % in a pointwise manner.
% \begin{equation}
%     L_\text{r} = \sum_{i,j}^{|Z^\text{r}|} \mathbb{1}_{y_i > y_j} * \log(1 + e^{s_j - s_i}),
% \end{equation}
% where $y_i$ represents the ranking label of $z_i$. Note that our DReranker only receives an input and a single demonstration, without including dependent demonstrations, which may not fully capture the dependency-aware ranking labels. Nonetheless, this design saves inference time, making our DReranker more efficient. We plan to explore architectures that can model multiple dependent demonstrations efficiently in the future.

\subsection{Inference}
During inference, we first encode the entire demonstration pool $\mathcal{P}$ using our trained DRetriever and build the index. Then, given a test input $I^\text{test}=(q^\text{test}, p^\text{test}_i)$, we first retrieve top-$D$ demonstrations using DRetriever. Then, we sequentially and greedily select the highest-scoring demonstration and append it to the ranking list. Specifically, the first demonstration is selected only conditioned on the test input $I^\text{test}$. Once the top-$i$ demonstrations have been selected, we choose the demonstration from the unselected ones that, when concatenated with the selected top-$i$ demonstration and $I^\text{test}$, results in the highest LLM score, and append it to the ranking list. This process is repeated until the ranking list contains $K$ demonstrations. Finally, we choose these $k$ demonstrations as the in-context demonstrations and concatenate them with the test input to calculate the relevance score. We perform this process for all retrieved passages of $q^\text{test}$ and rank these passages based on their relevance scores.

\section{Experiments}
\subsection{Setting}
\subsubsection{Datasets} In our experiments, we train and evaluate our DemoRank on diverse ranking datasets, including HotpotQA~\cite{hotpotqa}, NQ~\cite{NQ}, FEVER~\cite{fever} and MS MARCO~\cite{MSMARCO}. We use their training set to train our models respectively and evaluate the models on the corresponding test set (for MS MARCO, the evaluation is conducted on its development set as well as two in-domain datasets, TREC DL19~\cite{dl19} and TREC DL20~\cite{dl20}).

\subsubsection{Implementation Details} We use FLAN-T5-XL~\cite{flan} as the LLM scorer and passage ranker unless otherwise specified. For the construction of training input, we pair each query in the training set with one relevant passage and one irrelevant passage respectively, thus generating two training inputs. Next, we will introduce the training details of our models. As for DRetriever, we set the number of demonstration candidates $N$ as 50 and hyper-parameter $\lambda$ for multi-task learning as 0.2. Following previous study~\cite{llmr}, we use e5-base-v2~\cite{e5} to initialize DRetriever. As for DReranker, the number $M$ of retrieved demonstrations for scoring is 50 and the maximum iteration number $K$ in Section~\ref{subsec:reranker} is set as 3. We apply DeBERTa-v3-base~\cite{deberta} for model initialization. Both the DRetriever and DReranker are trained for 2 epochs. During inference, the number of retrieved demonstration $D$ for reranking is set as 30. 

\begin{table*}[t]
\centering
% \small
\setlength{\tabcolsep}{4mm}{
\caption{Main results (NDCG@10) on different datasets. The best results are marked in bold and the column Average represents the average performance of all datasets. "$\dagger$" indicates the model outperforms the best baseline significantly with paired t-test at $p$-value $<$ 0.05 level.}
\label{tab:main_exp}
\begin{tabular}{lcccccc|c}
\toprule
Method & HotpotQA & NQ & FEVER & DL19 & DL20 & MS MARCO & Average \\ \midrule
Initial Order & 63.30 & 30.55 & 65.13 & 50.58 & 47.96 & 22.84 & 46.73 \\
0-shot & 60.65 & 48.62 & 38.92 & 66.13 & 65.57 & 33.24 & 52.19 \\ \midrule
Random & 59.42 & 48.61 & 38.61 & 66.57 & 64.84 & 33.70 & 51.96 \\
K-means & 59.27 & 48.71 & 38.33 & 66.30 & 66.22 & 33.73 & 52.09 \\
DBS & 60.15 & 48.62 & 39.00 & 66.40 & 65.21 & 33.61 & 52.17 \\
BM25 & 63.18 & 49.78 & 40.19 & 66.08 & 65.85 & 34.03 & 53.19 \\
SBERT & 58.38 & 49.23 & 36.80 & 66.67 & 65.07 & 33.71 & 51.64 \\
E5 & 63.42 & 49.60 & 39.71 & 66.40 & 65.33 & 34.07 & 53.09 \\
DemoRank & \textbf{67.03}$^\dagger$ & \textbf{52.31}$^\dagger$ & \textbf{48.43}$^\dagger$ & \textbf{67.76} & \textbf{66.76} & \textbf{35.31}$^\dagger$ & \textbf{56.27} \\ \bottomrule
\end{tabular}}
\end{table*}

\subsubsection{Baselines} We compare our demonstration selection method with a series of baselines: 

\noindent$\bullet$ \textbf{Initial Order}: Following previous studies~\cite{rankgpt,beyond_yes_no}, we use the top-100 passages retrieved by BM25 as the initial passages order. All the subsequent baselines will perform reranking on these passages.

\noindent$\bullet$ \textbf{0-shot}: The passage ranking approach based on relevance generation without any demonstration.

\noindent$\bullet$ \textbf{Random}: We randomly sample demonstrations from the demonstration pool $\mathcal{P}$ for each test input.

\noindent$\bullet$ \textbf{DBS}~\cite{parade}: DBS is a rule-based selection approach based on query generation in passage ranking. It selects the demonstrations which are the most difficult for the LLM to predict. In this paper, we implemented the algorithm based on the relevance generation approach. We define a score for each demonstration as the probability of the LLM generating the corresponding relevance label given a query and passage. The demonstrations with the lowest scores are applied.

\noindent$\bullet$ \textbf{K-means}: K-means is another static demonstration selection approach. This method clusters all the demonstrations in the pool into $k$ clusters and then selects $k$ demonstrations closest to each cluster center for ICL. We use the E5~\cite{e5} model to obtain the demonstration embeddings for clustering.

\noindent$\bullet$ \textbf{BM25}~\cite{bm25}: BM25 is a widely-used sparse retriever. We apply it to retrieve demonstrations that are most similar to the test query.

\noindent$\bullet$ \textbf{SBERT}~\cite{sbert}: We use Sentence-BERT as the off-the-shelf demonstration retriever following~\cite{RubinHB22}\footnote{The checkpoint is from https://huggingface.co/sentence-transformers/paraphrase-mpnet-base-v2.}. We use SBERT to encode all the demonstrations in the pool and retrieve the most similar ones.

\noindent$\bullet$ \textbf{\textbf{E5}}~\cite{e5}: E5 is another off-the-shelf dense retriever. Following~\citet{llmr}, we use the same retrieval method as SBERT based on e5-base-v2 checkpoint\footnote{https://huggingface.co/intfloat/e5-base-v2}.

\subsection{Main Results}
We compare DemoRank with baselines in 3-shot ICL and the main results are shown in Table~\ref{tab:main_exp}. From the results, we draw the following observations: 

(1) Our framework DemoRank shows the best performance on all datasets and outperforms the second-best baselines with p-value $<$ 0.05 on most datasets, which indicates the significance of the improvements. For example, DemoRank outperforms the second-best model E5 on HotpotQA by about 4 points, and the second-best model BM25 on FEVER by about 8 points. The lack of significance on DL19 and DL20 is likely due to the small dataset size (only a few dozen queries), causing higher result variance. 

(2) Random selection and rule-based methods (such as K-means and DBS) struggle to achieve performance improvements and may even be harmful (\eg, on the HotpotQA dataset). This is quite different from the phenomenon observed in the NLP field~\cite{llmr, udr}, indicating that the demonstration selection for passage ranking is a very complex task.

(3) The similarity-based demonstration selection methods (\eg, BM25 and E5) perform well compared to other baselines. This suggests that the similarity between the demonstration and the input also helps select useful demonstrations. However, compared to DemoRank, they still fall significantly behind, proving that DemoRank can better model the relationships between the input and demonstrations, as well as the demonstration dependencies, thereby helping to select more effective few-shot demonstrations.

(4) The improvement from DemoRank compared to 0-shot varies greatly across datasets. For instance, DemoRank yields a 1-2 point gain on MSMARCO passage ranking datasets (MSMARCO, DL19, DL20) but nearly a 7-point boost on HotpotQA. We speculate that this is due to the varying difficulty levels of different datasets. HotpotQA is a multi-hop QA dataset, which requires more complex query understanding and reasoning than MSMARCO passage ranking datasets. Therefore, introducing demonstrations can better help the LLM understand the task, leading to greater improvements.

\subsection{Analysis}
In this section, we discuss different variants of DemoRank, compare DemoRank with different supervised passage rankers, and evaluate its performance using different demonstration numbers, unseen datasets, different LLM passage rankers and different demonstration retrieval numbers.

\subsubsection{Different Variants of DemoRank}
\label{subsubsec:variants}
To understand the effectiveness of each component in DemoRank, we further evaluate different variants of DemoRank. We conduct the experiments on FEVER, NQ and DL19 with 3-shot ICL, shown in Table~\ref{tab:variants}. First, we remove our demonstration reranker DReranker from DemoRank and only use demonstrations retrieved by our demonstration retriever DRetriever, denoted as ``- DemoRank w/o DReranker''. We can see that removing DReranker causes about 1.5 points drop, which proves the effectiveness of our DReranker. Secondly, to further validate the effectiveness of our dependency-aware reranking approach, we introduce another reranker variant that ignores the demonstration dependency, denoted as ``DemoRank w/o Dependency''. Specifically, given the demonstrations $Z_r$ retrieved by our DRetriever, this variant uses LLM to calculate a score of each candidate without considering other candidates, following the same scoring strategy as DRetriever. After that, the ranking list of scored demonstrations is used to fine-tune a demonstration reranker with the pairwise loss RankNet. Without considering the demonstration dependency, this variant lags behind DemoRank by about 2 points on FEVER and 1 point on average, which further proves that our dependency-aware reranking approach helps select more effective few-shot demonstrations.

We also investigate the effectiveness of ranking loss $L_r$ and the in-batch negatives on DRetriever training. Note that ``DemoRank w/o DReranker'' represents only applying DRetriever. Removing $L_r$ (denoted as DRetriever w/o $L_r$) causes about 0.6 points drop on average, which proving the effectiveness of ranking signals in scored demonstrations. Besides, incorporating the in-batch negatives into contrastive learning (denoted as DRetriever w/o IBN) brings no significant improvement and even causes a performance drop on DL19. This is because the utility of demonstrations in ranking tasks is not directly related to their similarity with the training input and the randomly sampled in-batch demonstrations may have a high score and act as a positive. Considering the lack of performance improvement and the additional computational overhead, we did not use in-batch negatives. 

Lastly, we also validate the training effectiveness of our DReranker on different demonstration retrievers by replacing our trained DRetriever with E5 in our whole framework, denoted as $\text{DemoRank}_\text{E5}$. From the results, we can see that $\text{DemoRank}_{\text{E5}}$ significantly improves E5, which proves that our DReranker's training is flexible and not restricted by specific demonstration retriever.

\begin{table}[]
\centering
% \small
\caption{Results (NDCG@10) of different variants.}
\label{tab:variants}
\setlength{\tabcolsep}{1.1mm}{
\begin{tabular}{lccc|c}
\toprule
Method & FEVER & NQ & DL19 & Average \\ \midrule
\multicolumn{4}{l|}{\textit{Ablation study}} & \multicolumn{1}{l}{} \\
- DRetriever w/ IBN & 44.43 & 51.68 & 67.14 & 54.42 \\
- DRetriever w/o $L_r$ & 43.65 & 50.65 & 67.65 & 53.98 \\
- DemoRank w/o DReranker & 44.40 & 51.69 & 67.70 & 54.60 \\
- DemoRank w/o Dependency & 46.64 & 51.72 & 67.12 & 55.16 \\
DemoRank & 48.43 & 52.31 & 67.76 & \textbf{56.17} \\ \midrule
\multicolumn{4}{l|}{\textit{Using E5 as demonstration retriever}} & \multicolumn{1}{l}{} \\
E5 & 39.71 & 49.60 & \multicolumn{1}{l|}{66.40} & 51.90 \\
$\text{DemoRank}_{\text{E5}}$ & 47.29 & 51.76 & \multicolumn{1}{l|}{68.19} & \textbf{55.75} \\ \bottomrule
\end{tabular}}
\end{table}

\begin{table}[]
\centering
\vspace{-1mm}
\caption{Results (NDCG@10) on MS MARCO, DL19 and DL20. QNum represents the number of queries used in the MS MARCO training set.}
\label{tab:supervised}
\begin{tabular}{llccc}
\toprule
QNum & Method & \multicolumn{1}{l}{MS MARCO} & \multicolumn{1}{l}{DL19} & \multicolumn{1}{l}{DL20} \\ \midrule
0 & 0-shot & 33.24 & 66.13 & 65.57 \\ \midrule
\multirow{4}{*}{500K} & monoBERT & 39.97 & 70.70 & 67.28 \\
 & monoT5 & \textbf{40.05} & 70.58 & \textbf{67.33} \\
 & monoFLAN-T5 & 36.22 & \textbf{70.72} & 66.26 \\
 & DemoRank & 35.31 & 67.76 & 66.76 \\ \midrule
\multirow{4}{*}{20K} & monoBERT & 30.69 & 63.61 & 59.32 \\
 & monoT5 & 29.79 & 61.16 & 52.72 \\
 & monoFLAN-T5 & 32.90 & 64.23 & 64.90 \\
 & DemoRank & \textbf{34.97} & \textbf{67.33} & \textbf{67.53} \\ \bottomrule
\end{tabular}
\end{table}

\subsubsection{Comparison with Supervised Rankers}
The training of DemoRank is primarily based on queries in the training set, which can also be used to directly finetune a supervised passage ranker. MonoBERT~\cite{monobert} and monoT5~\cite{monot5} are two supervised baselines widely used for comparison in passage ranking task. In this part, we choose monoBERT (340M) and monoT5 (220M) for comparison. Besides, we also introduce an LLM-based supervised ranker, named monoFLAN-T5, which uses the FLAN-T5-XL as the foundation model to enable a more fair comparison with DemoRank in LLM ranker size and adopts similar training strategy as monoT5. We compare DemoRank with the three baselines under different quantities of training queries (500K and 20K). We choose MS MARCO as the training set and NDCG@10 as the metric. We also report the 0-shot performance as a reference. The results are shown in Table~\ref{tab:supervised}. We can see that supervised models have a performance advantage when provided with 500K training queries. However, when the number of training queries is limited to 20K, DemoRank significantly outperforms them on three datasets and also shows a significant improvement over the 0-shot baseline. This suggests that when training data is limited, DemoRank is more effective than supervised models, highlighting the potential of DemoRank in low-resource scenarios.

% Training details of monoBERT and monoT5 are provided in Appendix~\ref{app:supervised_training}. We choose MS MARCO as the training set and NDCG@10 as the metric. We also report the 0-shot performance as a reference. The results are shown in Table~\ref{tab:supervised}. We can see that when provided with 500K queries, although DemoRank slightly outperforms monoBERT and monoT5 on DL20, it still lags behind them on DL19 and MS MARCO, indicating the advantages of supervised models when abundant training data is available. However, when the number of queries is limited to 20K, DemoRank significantly outperforms the two supervised models on three datasets and also shows a significant improvement over 0-shot baseline. This suggests that when training data is limited, DemoRank is more effective than supervised models, highlighting the potential of DemoRank in low-resource scenarios.

\begin{figure}[!t]
	\centering
        \vspace{1mm}
	\includegraphics[width=1\linewidth]{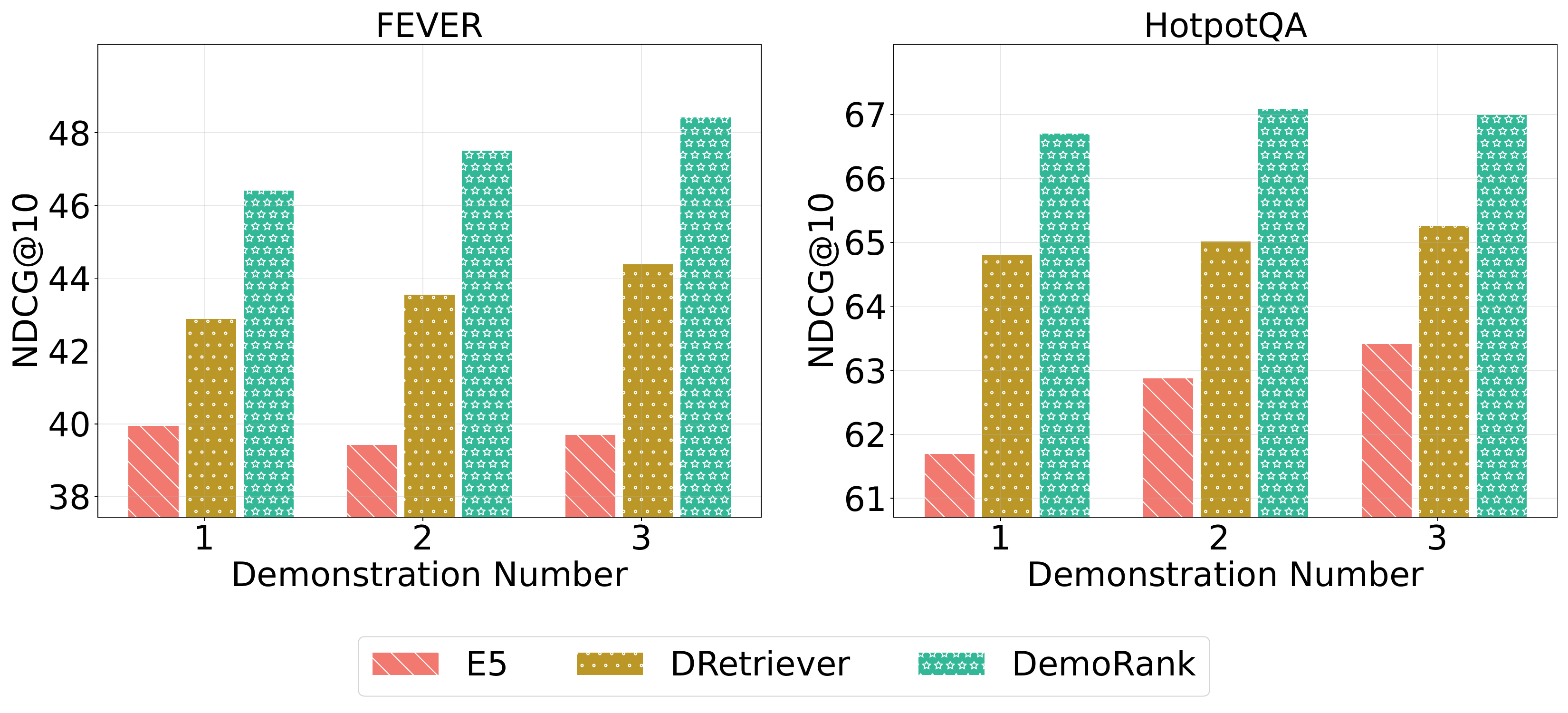}
	\caption{The impact of demonstration number.}
	\label{fig:length}
\end{figure}

\subsubsection{Different Demonstration Numbers}
Demonstration number is often considered a key factor affecting ICL. In this part, we discuss the performance of our models under different demonstration numbers. We compare DemoRank with E5 baseline on FEVER and HotpotQA datasets, using NDCG@10 as the metric. we also compare with our DRetriever to better show the improvement of our DReranker. The results are shown in Figure~\ref{fig:length}. We can see that both DRetriever and DemoRank outperform E5 consistently across different demonstration numbers, and our DReranker consistently improves DRetriever, which proves the effectiveness of our proposed models. Besides, we observe that when expanding from 2-shot to 3-shot, the performance of DemoRank on HotpotQA slightly decreases. We speculate that this is because 2-shot demonstrations already provide sufficient knowledge for the model, and adding more demonstrations may introduce unnecessary noise.

\begin{table*}[t]
\centering
\vspace{1mm}
\caption{Results (NDCG@10) on BEIR. Best results are marked in bold. We use MS MARCO's demonstration pool for retrieval and 3-shot ICL for E5 and DemoRank.}
\label{tab:beir_exp}
\setlength{\tabcolsep}{2.5mm}{
\begin{tabular}{lccccccc|l}
\toprule
Method & Robust04 & SCIDOCS & DBPedia & NEWS & FiQA & Quora & NFCorpus & \multicolumn{1}{c}{Avg} \\ \midrule
Initial Order & 40.70 & 14.90 & 31.80 & 39.52 & 23.61 & 78.86 & 33.75 & 37.59 \\ \midrule
MonoBERT & 44.18 & 15.99 & \textbf{41.70} & 44.62 & 32.06 & 74.65 & 34.97 & 41.17 \\
0-shot & 47.90 & 16.33 & 36.22 & 45.01 & 35.30 & 83.42 & 35.89 & 42.87 \\
E5 & 46.49 & 16.78 & 37.72 & 45.40 & 35.38 & \textbf{84.13} & 35.44 & 43.05 \\
DemoRank & \textbf{48.94} & \textbf{17.09} & 39.67 & \textbf{46.66} & \textbf{36.18} & 84.04 & \textbf{36.05} & \textbf{44.09} \\ \bottomrule
\end{tabular}}
\end{table*}

\subsubsection{Generalization on Unseen Datasets}
One of the application scenarios of DemoRank is its generalization on unseen datasets. To prove this, we evaluate DemoRank trained on MS MARCO dataset on a series of BEIR datasets. We choose 0-shot, E5 demonstration retriever, and the supervised passage ranker MonoBERT~\cite{monobert}, which is also trained on the MS MARCO dataset, for comparison. We use the demonstration pool from MS MARCO due to the lack of training sets in most BEIR datasets. The results are shown in Table~\ref{tab:beir_exp}. We can see that DemoRank outperforms the second-best model E5, by an average of about 1 point, proving its generalization ability. Furthermore, we also draw an interesting observation: despite using demonstrations from MS MARCO, DemoRank improves the 0-shot baseline across all datasets, indicating the potential of cross-dataset demonstrations in ICL.

\begin{table}[]
\centering
\caption{Results (NDCG@10) of different LLM ranker. We apply 3-shot ICL for BM25, E5 and DemoRank.}
\label{tab:different_ranker}
\setlength{\tabcolsep}{2.5mm}{
\begin{tabular}{lcccc}
\toprule
Method & FEVER & NQ & \multicolumn{1}{c|}{DL19} & Average \\ \midrule
Initial Order & 65.13 & 30.55 & \multicolumn{1}{c|}{50.58} & 46.73 \\ \midrule
\multicolumn{5}{c}{Flan-T5-XXL} \\ \midrule
0-shot & 37.38 & 47.61 & \multicolumn{1}{c|}{66.22} & 50.40 \\
BM25 & 43.89 & 50.47 & \multicolumn{1}{c|}{66.82} & 53.73 \\
E5 & 43.86 & 50.14 & \multicolumn{1}{c|}{66.45} & 53.48 \\
DemoRank & \textbf{50.89} & \textbf{51.68} & \multicolumn{1}{c|}{\textbf{69.23}} & \textbf{57.27} \\ \midrule
\multicolumn{5}{c}{Llama-3-8B-Instruct} \\ \midrule
0-shot & 27.53 & 36.24 & \multicolumn{1}{c|}{58.47} & 40.75 \\
BM25 & 29.22 & 35.31 & \multicolumn{1}{c|}{59.24} & 41.26 \\
E5 & 28.64 & 36.76 & \multicolumn{1}{c|}{57.36} & 40.92 \\
DemoRank & \textbf{45.50} & \textbf{37.11} & \multicolumn{1}{c|}{\textbf{60.93}} & \textbf{47.85} \\ \bottomrule
\end{tabular}}
\end{table}

\subsubsection{Transferability across different LLM Rankers}
\label{subsubsec:transferability}
In previous experiments, we used the same LLM (Flan-T5-XL) as the demonstration scorer and passage ranker. It is unknown whether the passage ranker could be replaced with other LLMs in the inference stage. In this section, we evaluate DemoRank's transferability across different LLM rankers on FEVER, NQ and DL19 datasets, and compare with several baselines including 0-shot, BM25, and E5. We experiment with Flan-T5-XXL (larger model size) and Llama-3-8B-Instruct (different model architecture) and the results are shown in Table~\ref{tab:different_ranker}. From the results, we can see that DemoRank outperforms all the baselines when using two different LLM rankers, proving its strong transferability. Besides, we observe that when using Flan-T5-XXL as LLM Ranker, DemoRank yields higher performance on FEVER and DL19 (50.89 and 69.23) compared with Flan-T5-XL (48.43 and 67.76 in Table~\ref{tab:main_exp}). This shows DemoRank's potential ability to improve passage ranking with larger-scale LLM rankers. 
% (3) Comparing the overall 0-shot performance between Flan-T5-XL (see Table~\ref{tab:main_exp}), Flan-T5-XXL and Llama-3-8B-Instruct, it is obvious that FlanT5 models perform better on average. This indicates that FlanT5 models are more suitable for passage ranking tasks, similar to findings from previous research~\cite{opensource}.

\begin{figure}[!t]
	\centering
	\includegraphics[width=1\linewidth]{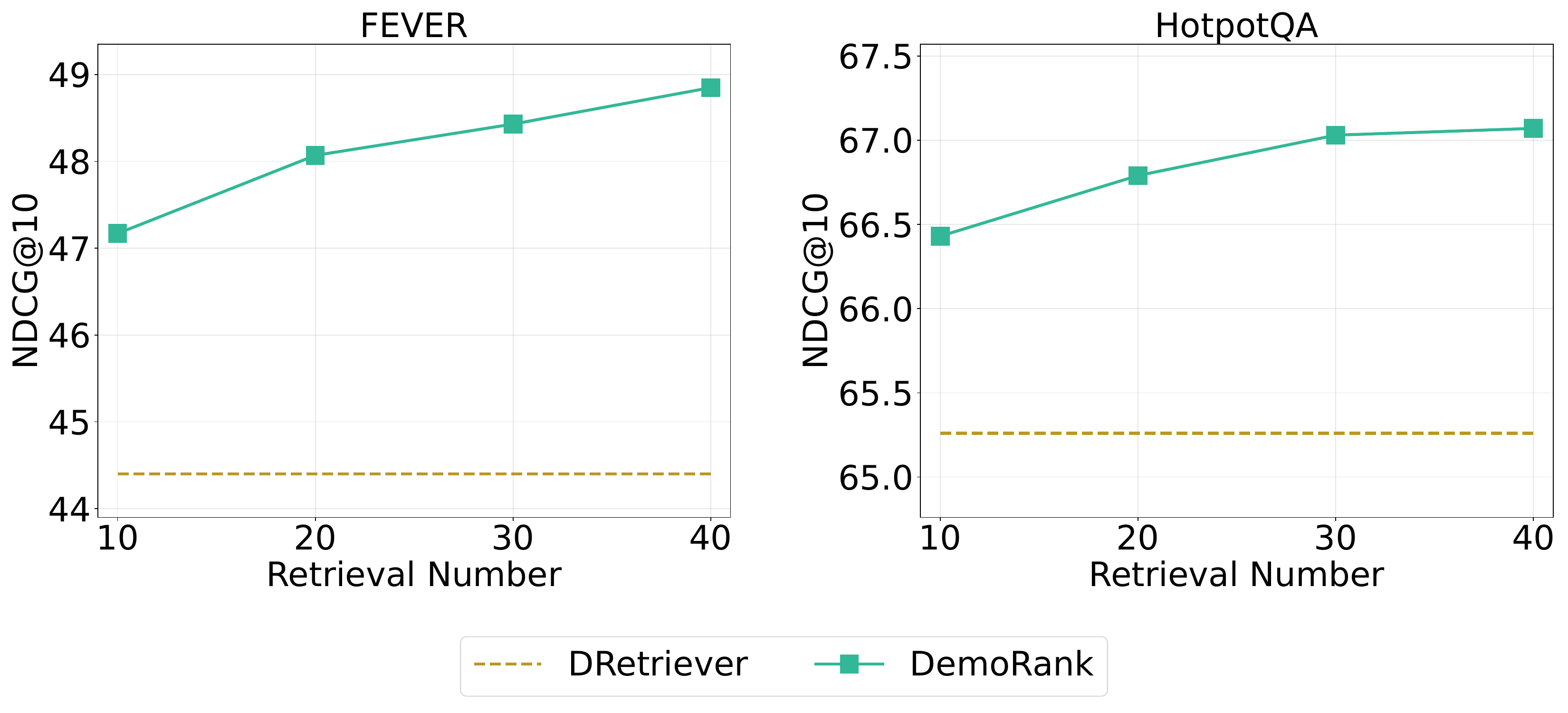}
	\caption{The impact of retrieved demonstration number on the performance of DemoRank.}
	\label{fig:retrieval_num}
\end{figure}

\subsubsection{The impact of retrieved demonstration number $D$}
Our DReranker aims to select few-shot demonstrations from top-$D$ demonstrations retrieved by DRetriever. Thus, DReranker's performance highly relies on the quality of these retrieved $D$ demonstrations. In this part, we intend to explore the impact of different retrieval numbers $D$ on the performance of DReranker. We choose different retrieval numbers $D$ (10, 20, 30 and 40), perform reranking using DReranker, and evaluate the performance of 3-shot ICL on FEVER and HotpotQA datasets. We also report the results of DRetriever as a reference. The results are shown in Figure~\ref{fig:retrieval_num}. From the results, we observe that increasing the retrieval number $D$ consistently enhances the DemoRank's performance on both datasets. This indicates that a larger pool of retrieved demonstrations allows the DReranker to select more effective few-shot demonstrations. Besides, DemoRank's performance on the HotpotQA dataset shows marginal improvement when the retrieval number increases from 30 to 40. This suggests that the most useful demonstrations are primarily concentrated within the top 30 retrieved ones on HotpotQA.

\section{Conclusion}
In this paper, we discuss the challenges of demonstration selection in ranking task and propose DemoRank framework which consists of a demonstration retriever and a dependency-aware reranker. Different from the demonstration retriever's training which models demonstrations independently, the demonstration reranker is trained using dependency-aware training samples which are constructed using an efficient method. We also design a novel list-pairwise training approach for reranker optimization, which compares a pair of demonstration lists that differ only in the last demonstration. Experiments on various ranking datasets prove the effectiveness of DemoRank. Further analysis shows the effectiveness of each proposed component, the advantages compared to supervised models, performance on different demonstration numbers, generalization on unseen datasets, etc.

% \section*{Limitations} \label{limitation}
% In this paper, we introduce a novel demonstration selection framework DemoRank for passage ranking task. We acknowledge several limitations in this paper that present opportunities for future work. First, due to limited computational resources, we can not conduct experiments with larger LLMs, such as those with 30B or even 70B parameters. Second, our framework is limited to pointwise passage ranking and lacks discussion on how demonstrations can be selected in pairwise and listwise passage ranking, which can be a promising direction to explore.

% \bibliographystyle{ACM-Reference-Format}
\balance
% \newpage
% \bibliography{references}

\end{document}